\documentclass[aps, reprint,showpacs,showkeys,pra,nofootinbib]{revtex4-1}
\usepackage{dcolumn}
\usepackage[utf8]{inputenc}
\usepackage{amsmath}
\usepackage{amsfonts}
\usepackage{amssymb}
\usepackage{graphicx}
\usepackage{hyperref}
\usepackage{xcolor}

\newcommand{\mb}{\mathbf}
\newcommand{\rme}{\mathrm{e}}

\begin{document}

\title{Local description of the Aharonov-Bohm effect with a quantum electromagnetic field}

\author{Pablo L. Saldanha}\email{saldanha@fisica.ufmg.br}
\affiliation{Departamento de F\'isica, Universidade Federal de Minas Gerais, Belo Horizonte, MG 31270-901, Brazil}

\date{\today}% It is always \today, today,

\begin{abstract}
In the seminal works from Santos and Gozalo [Europhys. Lett. \textbf{45}, 418 (1999)] and Marletto and Vedral [Phys. Rev. Lett.  \textbf{125}, 040401 (2020)], it is shown how the Aharonov-Bohm effect can be described as the result of an exchange of virtual photons between the solenoid and the quantum charged particle along its propagation through the interferometer, where both the particle and the solenoid interact locally with the quantum electromagnetic field. This interaction results in a local and gauge-independent phase generation for the particle propagation in each path of the interferometer. Here we improve the cited treatments by using the quantum electrodynamics formalism in the Lorentz gauge, with a manifestly gauge-independent Hamiltonian for the interaction and the presence of virtual longitudinal photons. Only with this more complete and gauge-independent treatment it is possible to justify the acquired phases for interferometers with arbitrary geometries, and this is an advantage of our treatment. We also extend the results to the electric version of the Aharonov-Bohm effect. Finally, we propose an experiment that could test the locality of the Aharonov-Bohm phase generation.
\end{abstract}

\maketitle

\section{Introduction}

In classical electrodynamics, the behavior of classical charged particles interacting with a classical electromagnetic field can be fully described in terms of local interactions between the particles and the gauge-independent electric and magnetic fields. However, in a quantum scenario the interference pattern formed by charged particles that propagate through an interferometer that encloses a long solenoid, like the one represented in Fig. 1, depends on the magnetic flux inside the solenoid even if the particles propagate only in regions where the electromagnetic fields are zero \cite{ehrenberg49,aharonov59}. This is the magnetic version of the Aharonov-Bohm (AB) effect \cite{aharonov59}, that was experimentally observed with many different systems \cite{chambers60,webb85,tonomura86,peshkin,bachtold99,peng10}. The original description of the AB effect involves a local interaction of the quantum particles with the gauge-dependent potential vector associated to the solenoid magnetic field. Other explanations for the AB effect relies on the superposition between the charged particles electromagnetic field and the solenoid electromagnetic field, that would affect the particles behavior in a nonlocal way \cite{liebowitz65,boyer73,boyer02,peshkin81,kang13,saldanha16}. In a novel interpretation, Vaidman deduced the AB phase using a quantum mechanical treatment for the charges of the solenoid interacting with the particles field \cite{vaidman12}, an idea that was further developed by Pearle and Rizzi \cite{pearl17a}. But in these works no local mechanism by which the charged particles acquire the AB phase is presented, such that a description of the effect in terms of local interactions between quantum particles and gauge-independent classical fields is not present so far. Since one of the principal ideas behind the introduction of the electromagnetic field concept is to avoid action at a distance, this fact may be disappointing. So there is still a vivid debate on locality issues of the AB effect \cite{kang13,saldanha16,vaidman12,pearl17a,pearl17b,aharonov15,vaidman15,aharonov16,kang17,marletto19}.

\begin{figure}\begin{center}
  \includegraphics[width=8cm]{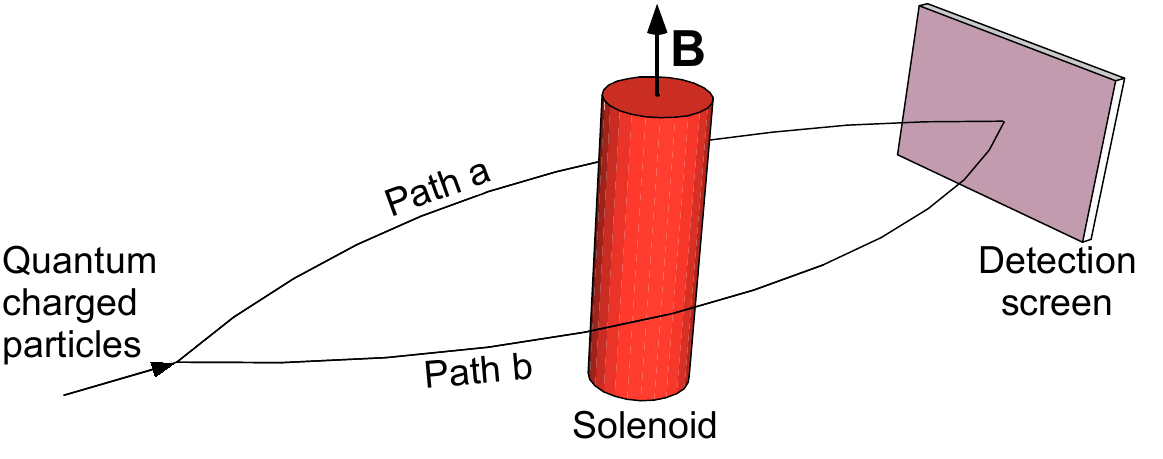}\\
  \caption{Magnetic version of the Aharonov-Bohm effect. The interference pattern of the quantum charged particles in the screen depends on the magnetic flux in the solenoid even if the particles only propagates in regions where the electromagnetic fields are zero.}
\label{fig1}
 \end{center}\end{figure}

In the seminal work from Santos and Gonzalo \cite{santos99}, the AB effect was treated with a quantum electromagnetic field. As emphasized by these authors, in quantum electrodynamics all interactions are local. Later other authors also used the quantized field to describe the AB effect in different situations \cite{pearl17b,marletto19,choi04,li18}. It is important to mention the recent contribution from Marletto and Vedral \cite{marletto19}, who generalized the results from Santos and Gonzalo, emphasized how the quantum particle locally acquires the AB phase difference along its propagation through a superposition of two different paths, and proposed experiments that could test the locality of the AB phase generation. In these treatments with a quantum electromagnetic field, the energy of the electromagnetic vacuum state depends on the particle path in the interferometer, due to an exchange of virtual photons between the particle and the solenoid, causing an extra phase difference between the paths (the AB phase). So, when the more fundamental quantum description of the electromagnetic field is used instead of the classical one, one of the the principal features of the electromagnetic field concept, which is to avoid action at a distance, is sustained. 

However, the treatments from Santos and Gonzalo \cite{santos99} and from Marletto and Vedral \cite{marletto19} considered the quantum electromagnetic field in the Coulomb gauge, without a manifestly covariant formulation. This fact makes their treatment valid only in specific interferometer configurations, with the quantum particle having a circular trajectory around a cylindrical solenoid. The treatments from Choi and Lee \cite{choi04} and from Pearle and Rizzi \cite{pearl17b} also used this specific interferometer configuration. In the work of Li, Hewak and Wang \cite{li18} a comparison between treatments of the AB effect using one-particle quantum mechanics and quantum field theory is presented, with the form of the potential vector operator to be used being deduced by imposing that both treatments must result in the same effect. Here we use a more fundamental form for the potential vector operator, without the need of further assumptions as in Ref. \cite{li18}. By using a covariant formulation in the Lorentz gauge, with the presence of virtual longitudinal photons, we generalize the results from refs. \cite{santos99,marletto19} for interferometers with arbitrary geometries. Also, we present a similar treatment in terms of a quantum electromagnetic field for the electric version of the AB effect \cite{aharonov59}. Finally, we present another experimental proposal that could verify the locality of the AB phase, different from the ones proposed in Ref. \cite{marletto19}, which we believe is easier to be implemented.

\section{Magnetic Aharonov-Bohm effect}

Consider the setup of Fig. 1, an interferometer for quantum charged particles with a solenoid enclosed by the two possible paths. The magnetic field $\mb{B}$ is confined inside the solenoid and the vector potential $\mb{A}$ is linked to the magnetic field through the relation $\mb{B}=\mb{\nabla}\times\mb{A}$. The particles can propagate only outside the solenoid, such that they experience no magnetic field and no Lorentz force, but are subjected to a nonzero vector potential. It can be shown that there is a contribution for the phase difference between the paths given by $q\Phi_0/\hbar$ (the AB phase), where $\Phi_0$ is the magnetic flux in the solenoid and $\hbar$ is Planck's constant divided by $2\pi$ \cite{aharonov59}. So the particles interference pattern depends on the enclosing magnetic flux even if they only propagate in regions with zero magnetic field. Some authors have argued that a longitudinal force could act on the particle causing lag between the electrons path that justifies the AB effect \cite{liebowitz65,boyer73,boyer02}, but experiments ruled out this hypothesis \cite{caprez07}. However, the presence of transverse forces in the AB scheme, as predicted in refs. \cite{shelankov98,berry99}, recently received an important indirect experimental confirmation by the measurement of the deflection of an electron beam by a magnetized nanorod \cite{becker19}.

Let us now treat the problem with a quantized electromagnetic field. To simplify the treatment, let us consider classical trajectories for each charged particle in the two possible paths represented in Fig.1. To justify this assumption, we can consider a particle wave function with reasonably well defined position and momentum along each path, forming a wave packet respecting the Heisenberg uncertainty relations. The momentum uncertainty must be much smaller than the average momentum value and the position uncertainty must be small enough such that spatial variations of the potential vector in this scale are negligible. On this way, we can compute the phase difference between the two paths due to the AB effect in a simple way, as we show below. 

The interaction of the quantum electromagnetic field with the solenoid can be represented by a term \cite{cohen}
\begin{equation}\label{v1}
	V_1=-\int d^3r'\mb{J}(\mb{r}')\cdot \mb{A}(\mb{r}')
\end{equation}
in the system Hamiltonian, where $\mb{J}$ is the (classical) current density of the solenoid and $\mb{A}$ the potential vector operator of the quantized electromagnetic field. The charge density in the solenoid is assumed to be zero. The potential vector operator can be written as \cite{cohen}
\begin{equation}\label{A}
	\mb{A}(\mb{r})=\int d^3k \sum_\sigma \sqrt{\frac{\hbar}{2\varepsilon_0\omega(2\pi)^3}}\;\hat{\epsilon}_{\mb{k}\sigma}a_\sigma(\mb{k})\rme^{i\mb{k}\cdot\mb{r}}+\mathrm{H.c.},
\end{equation}
where $\hat{\epsilon}_{\mb{k}\sigma}$ is an unitary polarization vector, $a_\sigma(\mb{k})$ is the annihilation operator for a mode with wavevector $\mb{k}$ and polarization index $\sigma$, and $\omega=ck$ is the mode angular frequency. In the traditional quantum optics treatments in the Coulomb gauge \cite{mandel}, the polarization index $\sigma$ assumes two values for each wavevector $\mb{k}$, corresponding to two polarizations orthogonal to $\mb{k}$. However, in the complete quantum electrodynamics formalism constructed in the Lorentz gauge \cite{cohen}, there are also photon modes with longitudinal polarization in the direction $\mb{\hat{k}}$, such that we have 3 indexes for $\sigma$ for each value of $\mb{k}$. Although there can be no \textit{real} photons with longitudinal polarizations, the presence of \textit{virtual} photons with longitudinal polarization is important in the present treatment.

The interaction of the electromagnetic field with the charged particle can be represented by a term \cite{cohen}
\begin{equation}\label{v2}
	V_2=-\frac{q}{m}\mb{p}\cdot \mb{A}(\mb{r})+qU(\mb{r})
\end{equation}
in the system Hamiltonian, where $q$, $m$, $\mb{p}$ and $\mb{r}$ are the particle charge, mass, momentum and position respectively, and $U$ is the scalar potential operator of the quantized electromagnetic field.  This operator can be written as \cite{cohen}
\begin{equation}\label{A0}
	U(\mb{r})=c\int d^3k \sqrt{\frac{\hbar}{2\varepsilon_0\omega(2\pi)^3}}\big[a_0(\mb{k})\rme^{i\mb{k}\cdot\mb{r}}+ \bar{a}_0(\mb{k})\rme^{-i\mb{k}\cdot\mb{r}} \big],
\end{equation}  
where $a_0(\mb{k})$ is the annihilation operator for a scalar photon mode with wavevector $\mb{k}$ and $\bar{a}_0(\mb{k})=-a^\dag_0(\mb{k})$. The definition of $\bar{a}_0(\mb{k})$ and its use in the above equation is necessary for having a consistent norm for the scalar photon states \cite{cohen}. Although there can be no \textit{real} scalar photons, there may be \textit{virtual} scalar photons. These are not relevant for the present problem, since the solenoid is assumed to be free of charge densities and so do not couple to scalar photons, but virtual scalar photons are used, for instance, to deduce the Coulomb law from quantum electrodynamics \cite{cohen}. Virtual scalar photons are also relevant in the electric version of the AB effect, to be treated later.

The idea of our treatment is to compute the change on the energy of the vacuum state of the electromagnetic field perturbed by the terms of Eqs. (\ref{v1}) and (\ref{v2}) in the system Hamiltonian, with the use of second-order perturbation theory. We show that the perturbed energy depends on the particle path in Fig. 1 and compute the phase difference for the particle propagation in the two possible paths. 

The terms of Eqs. (\ref{v1}) and (\ref{v2}) change the energy of the unperturbed electromagnetic vacuum state $|\mathrm{vac}\rangle$. Since all terms of $V_1$ and $V_2$ have annihilation or creation operators due to the presence of the operators $\mb{A}$ from Eq. (\ref{A}) and $U$ from Eq. (\ref{A0}), the first-order correction of the electromagnetic vacuum energy is zero. The second-order correction can be written as \cite{cohen2}
\begin{equation}
	\Delta E= \sum_{n\neq0}\sum_i\frac{|\langle \phi_n^i|(V_1+V_2)|\mathrm{vac}\rangle|^2}{E_0^0-E_n^0},
\end{equation}
where $E_0^0$ is the energy of the unperturbed vacuum, $|\phi_n^i\rangle$ represent eigenstates of the unperturbed Hamiltonian with energies $E_n^0$, and the index $i$ is associated to the degeneracy of these eigenstates (the vacuum state is nondegenerate). Terms with the product of matrix elements of $V_1$ with $V_1$ and of $V_2$ with $V_2$ result in self-energies of the solenoid and  of the particle respectively, not being relevant for the present study. Considering only the correction in energy due to the interaction between the solenoid and the particle mediated by the field, note that only field states with one photon are relevant, that can be represented as $|\mb{k},\sigma\rangle$, $\mb{k}$ being the wavevector and $\sigma$ the polarization index of the photon (scalar photons do not contribute). So we can write the part of the vacuum energy change which is relevant for the present study as
\begin{equation}\label{deltaE}
	\Delta E'= 2\mathrm{Re}\Bigg[\int d^3k\sum_\sigma\frac{\langle\mathrm{vac}|V_2|\mb{k},\sigma\rangle\langle \mb{k},\sigma|V_1|\mathrm{vac}\rangle}{-\hbar\omega}\Bigg],
\end{equation}
$\hbar\omega=\hbar ck$ being the photon energy.  Using Eqs. (\ref{v1}) and (\ref{A}), we obtain
\begin{equation}
	\langle \mb{k},\sigma|V_1|\mathrm{vac}\rangle = -\int d^3r' \;\sqrt{\frac{\hbar}{2\varepsilon_0\omega(2\pi)^3}}\;\mb{J}(\mb{r}')\cdot\hat{\epsilon}_{\mb{k}\sigma}\;\rme^{-i\mb{k}\cdot\mb{r}'}.
\end{equation}
Using Eqs. (\ref{v2}), (\ref{A}), and (\ref{A0}), we obtain
\begin{equation}
	\langle\mathrm{vac} |V_2|\mb{k},\sigma\rangle =  -\frac{q}{m}\sqrt{\frac{\hbar}{2\varepsilon_0\omega(2\pi)^3}}\;\mb{p}\cdot\hat{\epsilon}_{\mb{k}\sigma}\;\rme^{i\mb{k}\cdot\mb{r}}.
\end{equation}
So Eq. (\ref{deltaE}) becomes
\begin{equation}\label{deltaEl}
	\Delta E'=\frac{-q}{mc^2\varepsilon_0}\mathrm{Re}\Bigg\{\int d^3r'\Bigg[\int d^3k \frac{\rme^{i\mb{k}\cdot(\mb{r}-\mb{r}')}}{(2\pi)^3k^2}\Bigg]\mb{J}(\mb{r}')\cdot\mb{p}\Bigg\}.
\end{equation}
Note that if we consider only two orthogonal polarization vectors $\hat{\epsilon}_{\mb{k}\sigma}$ for each wavevector $\mb{k}$, as in the treatments of Refs. \cite{santos99,marletto19} in the Coulomb gauge without longitudinal photons, Eq. (\ref{deltaE}) reduces to Eq. (\ref{deltaEl}) only in specific configurations, as with a cylindrical solenoid and circular paths for the quantum particle. Our treatment with the presence of virtual longitudinal photons generalize this result for arbitrary configurations. The term inside brackets in Eq. (\ref{deltaEl}) can be written as 
\begin{equation}
	\int d^3k \frac{\rme^{i\mb{k}\cdot(\mb{r}-\mb{r}')}}{(2\pi)^3k^2} = \frac{1}{4\pi|\mb{r}-\mb{r}'|}.
\end{equation}
So we can write
\begin{equation}\label{deltaEE}
	\Delta E'=-\frac{q}{m}\mb{p}\cdot\mb{\mathcal{A}}(\mb{r}),
\end{equation}
where
\begin{equation}\label{aa}
	\mb{\mathcal{A}}(\mb{r})=\frac{\mu_0}{4\pi}\int d^3r' \frac{\mb{J}(\mb{r}')}{|\mb{r}-\mb{r}'|}
\end{equation}
is the potential vector generated by the current in the solenoid in the Lorentz gauge \cite{jackson}.

The phase difference between the two paths of Fig. 1 can then be written as 
\begin{equation}\label{phi}
	\varphi = -\int \frac{(\Delta E'_b-\Delta E'_a)}{\hbar}dt, 
\end{equation}
where $\Delta E'_a$ and $\Delta E'_b$ are the corrections in energy on paths $a$ and $b$ respectively, given by Eq. (\ref{deltaEE}). By writing $\mb{p}/m = d\mb{l}/dt$, where  $d\mb{l}$ is an infinitesimal displacement around the path, we can write
\begin{equation}\label{AB phase}
	\varphi = \frac{q}{\hbar}\int_{\mathrm{path \;b}} \mathcal{A}\cdot d\mb{l} -\frac{q}{\hbar}\int_{\mathrm{path \;a}} \mathcal{A}\cdot d\mb{l} =\frac{q}{\hbar}\oint \mathcal{A}\cdot d\mb{l} = \frac{q\Phi_0}{\hbar},
\end{equation}
the AB phase.

The treatment we used here to arrive at Eq. (\ref{deltaEE}) is analogous to the deduction of the Coulomb interaction between two charges from similar changes on the vacuum electromagnetic energy due to the coupling of the quantized electromagnetic field with the charges \cite{cohen}. A difference is that the scalar photons are the relevant ones in this case. The usual interpretation is that the Coulomb attraction or repulsion is the result of the exchange of virtual photons between the charges \cite{cohen}. A similar interpretation can be used here. The energy variation obtained from  Eq. (\ref{deltaE}) has the contribution from terms with a creation operator in $V_1$ and a annihilation operator in $V_2$ for the same photon mode, that can be interpreted as the result of a virtual photon emission by the solenoid which is absorbed by the quantum particle. In an analogous way, terms with a annihilation operator in $V_1$ and a creation operator in $V_2$ for the same photon mode can be interpreted as the result of a virtual photon emission by the quantum particle which is absorbed by the solenoid.  So we interpret the AB effect in the quantum electrodynamics formalism as the result of an interaction between the quantum particle and the solenoid mediated by the exchange of virtual photons, where both the solenoid and the particle interact locally with the quantum electromagnetic field. In the AB scheme of Fig. 1, there is entanglement between the particle path and the energy of the vacuum state of the electromagnetic field, which is the responsible for the appearance of the AB phase.

An interesting point is that the quantity $\mb{\mathcal{A}}$ that appears in Eq. (\ref{deltaEE}) is a gauge-independent quantity defined by the expression (\ref{aa}), which coincides with the potential vector generated by the solenoid current density in the Lorentz gauge. This is the case because the quantity $\Delta E'$ from Eq. (\ref{deltaEE}) cannot be gauge-dependent, since it represents a physical energy variation in the system. So a well defined gauge-independent phase is acquired by each part of the particle wave function while it is propagating through each path. In the following we discuss a possible experiment that could verify if the AB phase is indeed locally generated.

\section{Experimental proposal}

Consider the situation depicted in Fig. 2. A source releases an ion that falls in the $-\mb{\hat{y}}$ direction due to gravity. Its wave function spreads in the process, and while it is falling the ion traps A and B are turned on. If the ion is not detected on the detection screen, its wave function will be coherently superposed in traps A and B. If the traps are now turned off, the ion wave function evolves falling again until the ion is detected on the screen. By repeating the experiment many times, an interference pattern is formed. Since the solenoid is not enclosed by the paths, the interference pattern does not depend on its magnetic flux. 

\begin{figure}\begin{center}
  \includegraphics[width=8cm]{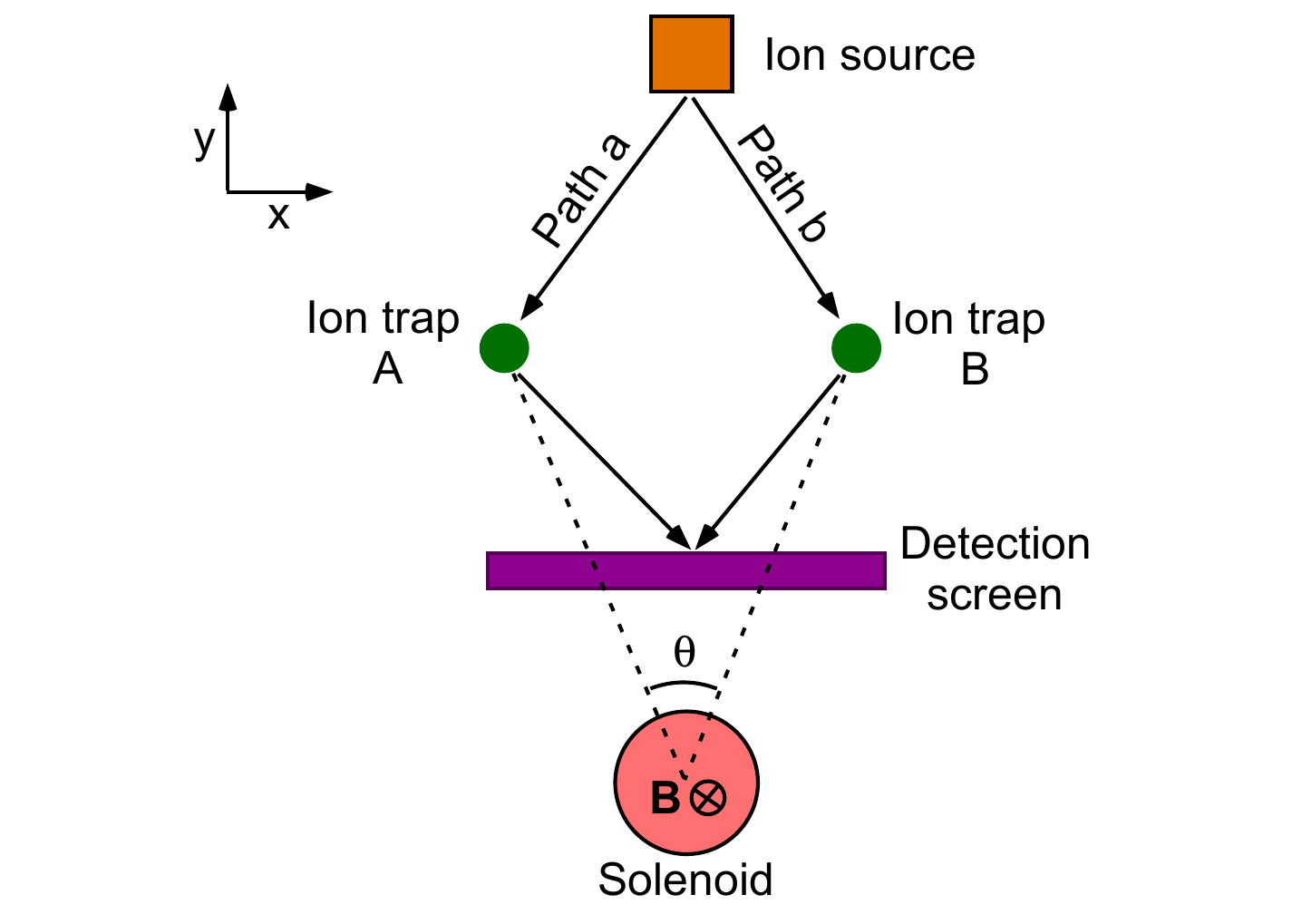}\\
  \caption{Scheme to verify the locality of the Aharonov-Bohm phase generation. An ion is released from the source and evolves to a superposition of being trapped at ion traps A and B. The magnetic field of the solenoid is then turned off. When the ion is released from the traps, the interference pattern on the detection screen should depend on an intermediate Aharonov-Bohm phase, even if the particle paths do not enclose the solenoid.}
\label{fig2}
 \end{center}\end{figure}

Consider now a situation where, while the ion wave function is superposed in the two traps A and B, the magnetic field of the solenoid is turned off. By Faraday's law, an electric field will then be produced in the position of the traps A and B. But if the solenoid has a cylindrical symmetry and the traps are positioned at the same distance from it, the magnitude of the electric field at each trap will be the same. Considering also that the traps are identical and have a symmetric potential in the $x-y$ plane, this induced electric field (considered to be small) will change the energy of the portion of the ion wave function in each trap in a perturbative way equally, producing no phase difference in the process. If, after the magnetic field of the solenoid is reduced to zero, the traps are turned off, the expected interference pattern on the screen should depend on an intermediate AB phase. This is because there is an accumulated phase difference while the ion goes from the source to traps A and B, but no extra phase when the ion propagates from the traps to the screen, since the solenoid current is zero in this case. Using cylindrical coordinates $(r,\phi,z)$ with the origin at the solenoid axis, the potential vector in the Lorentz gauge with a magnetic flux $\Phi_0$ in a long cylindrical solenoid is given by $\mathcal{A}=-\Phi_0/(2\pi r)\mb{\hat{\phi}}$ \cite{Afanasiev}. Using Eqs. (\ref{deltaEE}) and (\ref{phi}) and following the same steps that led us to Eq. (\ref{AB phase}), we deduce an AB phase
\begin{equation}\label{AB phase 2}
	\varphi' = \frac{q}{\hbar}\int_{\mathrm{path \;b}} \mathcal{A}\cdot d\mb{l} -\frac{q}{\hbar}\int_{\mathrm{path \;a}} \mathcal{A}\cdot d\mb{l} = \frac{\theta}{2\pi} \frac{q\Phi_0}{\hbar}
\end{equation}
in this experimental situation, with $\theta$ defined in Fig. 2. An experiment like this should then show an intermediate AB phase even if the particle paths do not enclose the solenoid, demonstrating that that the AB phase is locally generated as predicted by the present treatment.

\section{Electric Aharonov-Bohm effect}

Now let us discuss the electric version of the AB effect \cite{aharonov59}, represented in Fig. 3. The experiment should be made with quantum charged particles sent one by one to the interferometer. The scalar potentials of the conductor tubes in each path are zero when the particle is outside them. When the quantum particle is in a superposition of being inside each of the tubes, the potential of the tubes is varied and comes back to zero before the particle wave function exits each of them. In this case the particles experience no electric field and no Lorentz force, but experience different electric potentials in each path. Again, this results in a phase difference between the paths. If the potential of the tube of path $a$ is $U_a(t)$ and the one of path $b$  is $U_b(t)$,  it can be shown that the phase difference is $\int dt[U_a(t)-U_b(t)]q/\hbar$, affecting the interference pattern \cite{aharonov59}. This behavior was also experimentally verified \cite{matteucci85,vanou98}, although not with the charged particles propagating only in free-field regions.

\begin{figure}\begin{center}
  \includegraphics[width=8cm]{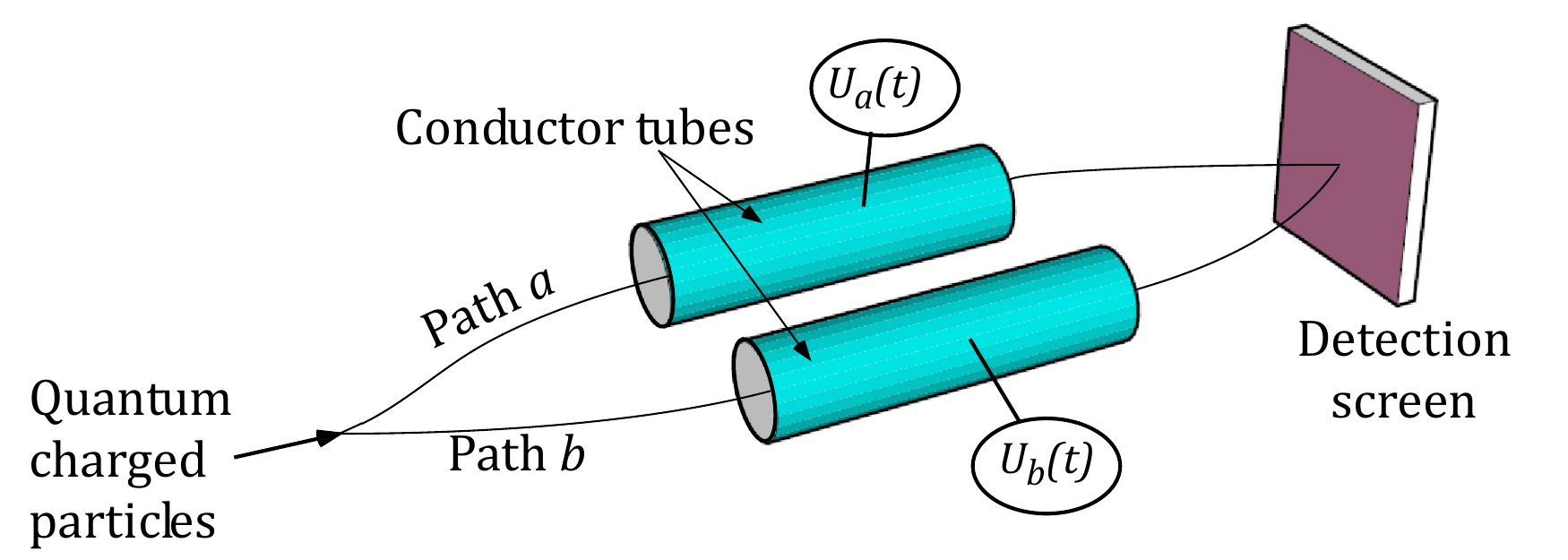}\\
  \caption{Electric version of the Aharonov-Bohm effect. The interference pattern of the charged quantum particles in the screen depends on the potentials of the conductor tubes even if the particles only propagate in regions where the electromagnetic fields are zero, with the tubes potentials being set to zero when the particles wave function is outside them.}
\label{fig3}
 \end{center}\end{figure}

The treatment of the electric version of the AB effect with a quantized electromagnetic field is completely analogous to the magnetic AB effect treated previously. But now the term in the Hamiltonian that represents the interaction of the quantum electromagnetic field with the charges on the tubes is
\begin{equation}
	V_1=\int d^3r'\rho(\mb{r}')U(\mb{r}'),
\end{equation}
where $\rho(\mb{r}')$ is the charge density of the tubes and $U(\mb{r}')$ is the scalar potential operator given by Eq. (\ref{A0}). The current density in the tubes is assumed to be zero. The interaction of the charged particle with the electromagnetic field is still given by Eq. (\ref{v2}). Following the same steps as before, it is straightforward to show that the change on the electromagnetic vacuum energy when the particle is in a position $\mb{r}$, disregarding self-energy terms, is given by
\begin{equation}
	\Delta E'=q\,\mathcal{U}(\mb{r})\;\; \mathrm{with}\;\;\mathcal{U}(\mb{r})=\int d^3r'\frac{\rho(\mb{r}')}{4\pi\varepsilon_0|\mb{r}-\mb{r}'|}.
\end{equation}
It is straightforward to show that this result leads to the predicted phase difference between the paths in the electric version of the AB effect.

\section{Conclusion}

To summarize, we have generalized the treatments of Refs. \cite{santos99,marletto19} to show that the AB effect in the scheme depicted in Fig. 1 (Fig. 3) can be described as the result of an interaction between the charged particles and the solenoid (conductor tubes) mediated by the quantum electromagnetic field in a local way, with the exchange of virtual photons. The particles locally acquire an AB phase due to the change of the energy of the vacuum electromagnetic field, which depends on the particle path. Our treatment also predicts that intermediate AB phases can be measured even if the quantum particles possible paths do not enclose the solenoid, what would demonstrate that the AB phase is indeed locally generated.

\begin{acknowledgements}

The author acknowledges Chiara Marletto, Vlatko Vedral, and Ana J\'ulia Mizher for very useful discussions. This work was supported by the Brazilian agencies CNPq, CAPES, and FAPEMIG. 

\end{acknowledgements}

%\acknowledgments

%This work was supported by the Brazilian agencies CNPq, CAPES and FAPEMIG.

%\bibliography{ref}

\end{document}